\documentclass[aps,prl,amsmath,amssymb,preprint,floatfix,lengthcheck,superscriptaddress]{revtex4-1}

\usepackage{graphicx}
\usepackage[centering,hmargin=20mm,tmargin=30mm,bmargin=25mm]{geometry}
\usepackage{multirow}
\usepackage{newtxtext}
\usepackage[cmintegrals]{newtxmath}
\usepackage{bm}
\usepackage{bbm}
\usepackage{etoolbox}
\usepackage{setspace}
\usepackage{subfigure}
\usepackage{enumitem}
\usepackage{titlesec}

\usepackage{xcolor}
\usepackage{hyperref}
\usepackage{cleveref}
\hypersetup{colorlinks,
	linkcolor={blue!75!black!80!yellow},
	citecolor={blue!75!black!80!yellow},
	urlcolor={blue!75!black!80!yellow}
}

\makeatletter
\renewcommand\@make@capt@title[2]{	\@ifx@empty\float@link{\@firstofone}{\expandafter\href\expandafter{\float@link}}	\sffamily{\textbf{#1}}\@caption@fignum@sep#2
}
\makeatother

\crefname{Fig}{Fig.}{Figs.}
\Crefname{Fig}{Figure.}{Figures.}

\begin{document}

\title{Non-Equilibrium Phonon Transport Across Nanoscale Interfaces}
\author{Georgios Varnavides}
\affiliation{Department of Materials Science and Engineering, Massachusetts Institute of Technology, Cambridge, MA, USA}
\affiliation{Research Laboratory of Electronics, Massachusetts Institute of Technology, Cambridge, MA, USA}
\affiliation{John A. Paulson School of Engineering and Applied Sciences, Harvard University, Cambridge, MA,USA}
\author{Adam S. Jermyn}
\affiliation{Kavli Institute for Theoretical Physics, University of California at Santa Barbara, Santa Barbara, CA 93106, USA}
\affiliation{Institute of Astronomy, University of Cambridge, Madingley Rd, Cambridge CB3 0HA, UK}
\author{Polina Anikeeva}
	\email[Electronic address:\;]{anikeeva@mit.edu}
\affiliation{Department of Materials Science and Engineering, Massachusetts Institute of Technology, Cambridge, MA, USA}
\affiliation{Research Laboratory of Electronics, Massachusetts Institute of Technology, Cambridge, MA, USA}
\author{Prineha Narang}
	\email[Electronic address:\;]{prineha@seas.harvard.edu}
\affiliation{John A. Paulson School of Engineering and Applied Sciences, Harvard University, Cambridge, MA,USA}
\date{\today}

\begin{abstract}
Despite the ubiquity of applications of heat transport across nanoscale interfaces, including integrated circuits, thermoelectrics, and nanotheranostics, an accurate description of phonon transport in these systems remains elusive.
Here we present a theoretical and computational framework to describe phonon transport with position, momentum and scattering event resolution. 
We apply this framework to a single material spherical nanoparticle for which the multidimensional resolution offers insight into the physical origin of phonon thermalization, and length-scale dependent anisotropy of steady-state phonon distributions.
We extend the formalism to handle interfaces explicitly and investigate the specific case of semi-coherent materials interfaces by computing the coupling between phonons and interfacial strain resulting from a periodic array of misfit dislocations. 
Our framework quantitatively describes the thermal interface resistance within the technologically relevant Si-Ge heterostructures. 
In future, this formalism could provide new insight into coherent and driven phonon effects in nanoscale materials increasingly accessible via ultrafast, THz and near-field spectroscopies.
\end{abstract}
\maketitle

Understanding phonon-mediated heat transfer at the nanoscale is essential to the design and optimization of heat management for a variety of engineering systems including thermoelectrics~\cite{2011_Toberer}, nanoelectronics~\cite{1986_Walle,2006_Adhikari}, catalytic cells~\cite{2017_Robatjazi}, and nanotheranostics~\cite{2012_Ho}.
Advances in ultrafast probes of coherent dynamics have revealed non-equilibrium regimes of phonon transport, necessitating a new theoretical framework describing these effects~\cite{2017_Ishioka,2016_Ishioka,2011_Maznev,2003_Sokolowski-Tinten}.

The phenomenological heat conduction equation, can be deduced within the formalism of the Boltzmann Transport Equation (BTE) in the hydrodynamic limit~\cite{2016_Peraud}, but is known to breakdown at both short length- and time-scales~\cite{2009_Siemens,2011_Minnich,2013_Johnson,2013_Regner}, as well as in low-dimensional materials~\cite{2008_Chang}.
Conversely, one microscopic description of lattice thermal transport is the phonon BTE (pBTE), first formulated by Peierls in 1929~\cite{1929_Peierls}. 
This formalism provides the most-general description of semi-classical phonon transport by tracking the evolution of probability distributions in full phase space, resolving both spatial and momentum degrees of freedom. 
Shortly after it was proposed, linearized solutions of the pBTE enabled predictions of lattice thermal conductivity of crystalline insulators using the relaxation time approximation~\cite{1958_Klemens}, which assumes all perturbations return to equilibrium with the same timescale~\cite{1959_Callaway}. 
Recent computational methods have enabled the linearized pBTE to be solved exactly using materials parameters determined from first principles via variational~\cite{2013_Fugallo,1969_Hamilton,1976_Srivastava}, iterative~\cite{2005_Broido,1996_Omini}, and direct approaches~\cite{2016_Cepellotti,2013_Chaput,1966_Guyer}. 
Each of these theoretical studies solved the time and space-independent form of the pBTE, i.e. at a steady state assuming a spatially homogeneous structure, reducing the problem dimensionality to the three momentum degrees of freedom. 
An accurate picture of heat transport addressing finite size-effects and transport across interfaces, however, must include the spatial degrees of freedom of the nanoscale geometry~\cite{2015_Romano}. 
This \textit{Letter} aims to fill this void in theoretical transport methods, by incorporating all momentum degrees of freedom into a spatially-resolved BTE solver, building on our previous work in excited carrier dynamics~\cite{2017_Jermyn,2016_Brown,2017_Brown}. 

The starting point for describing steady-state phonon transport is the time-independent BTE given by:
\begin{equation}
\boldsymbol{v}_{\boldsymbol{q},s}\cdot \nabla n(\boldsymbol{q},s,\boldsymbol{r}) = G_0(\boldsymbol{q},s,\boldsymbol{r})+\Gamma_{\boldsymbol{q},s}\left[n\right], \tag{1} \label{eq:1}
\end{equation}
\noindent
where $\boldsymbol{q}$ and $s$ are the phonon momentum and polarization respectively, $\boldsymbol{r}$ is position, $n(\boldsymbol{q},s,\boldsymbol{r})$ is the phonon distribution function, and $\boldsymbol{v}_{\boldsymbol{q},s}$ is the phonon group velocity.
The term on the left reflects phonon drift with velocity $\boldsymbol{v}_{\boldsymbol{q},s}$, while the terms on the right account for phonon generation ($G_0(\boldsymbol{q},s,\boldsymbol{r})$), and collisions ($\Gamma_{\boldsymbol{q},s}\left[n\right]$).
\begin{figure*}[ht]
\includegraphics[width=\linewidth]{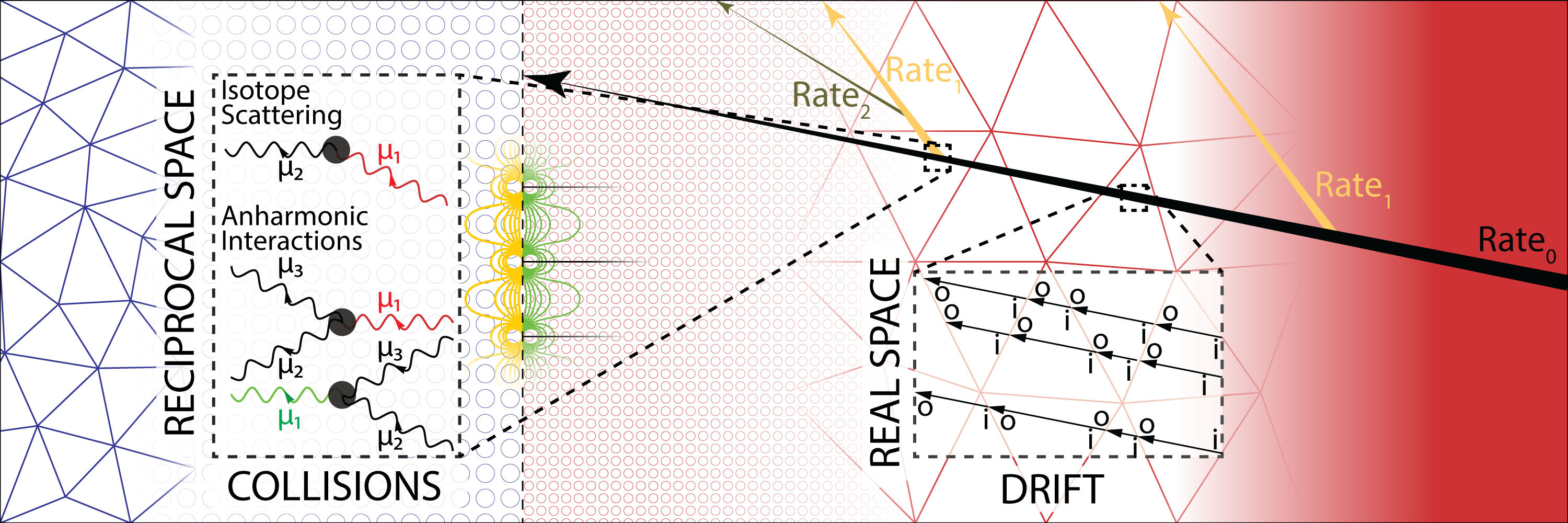}
\caption{Schematic description of the recursive formulation and computational implementation of the pBTE. 
Initially, phonons are injected at a constant rate, $G_0$, and allowed to drift through the real-space structure in accordance with their group velocity and mean-free path.
We solve the differential~\cref{eq:5a} using the finite element method with linear elements.
The mixing matrix, consisting of three-phonon interactions and isotopic scattering, is then applied to the unscattered distribution $\psi_0$ according to~\cref{eq:5b} to obtain the rate at which phonons are generated due to the first scattering event, $G_1$.
The red and blue colors illustrate `hot' and `cold' sides of the material respectively, while the green and yellow dilatational contours illustrate that the formalism is capable of describing strained interfaces.}
\label[Fig]{fig:1}
\end{figure*}

Using a series of approximations detailed in~\cite{2017_Jermyn,1990_Srivastava}, we linearize~\cref{eq:1} in a steady state to return:
\begin{equation}
\boldsymbol{v}_\mu \cdot \nabla \psi(\mu,\boldsymbol{r}) = G_0(\mu,\boldsymbol{r}) +\sum_{\mu'} A_{\mu\mu'}\psi(\mu',\boldsymbol{r}), \tag{2} \label{eq:2}
\end{equation}
\noindent
where $\mu$ is a combined phonon label encapsulating $\boldsymbol{q}$ and $s$.
The scattering matrix $A_{\mu\mu'}$ specifies the rate at which phonons scatter from state $\mu$ into state $\mu'$.
This arises from the first order term in the Taylor expansion of $\Gamma_{\boldsymbol{q},s}\left[n\right]$ around $\bar{n}(\mu,\boldsymbol{r})$, the equilibrium Bose-Einstein distribution.
Deviation, $\psi(\mu,\boldsymbol{r})$ away from equilibrium is defined via
\begin{align}
n(\mu,\boldsymbol{r}) &\approx \bar{n}(\mu,\boldsymbol{r}) - k_B T \psi(\mu,\boldsymbol{r})\frac{\partial \bar{n}(\mu,\boldsymbol{r})}{\partial \left( \hbar \omega_{\mu} \right)} \tag{3a} \label{eq:3a} \\
&= \bar{n}_\mu+\psi_{\mu} \bar{n}_{\mu}(\bar{n}_{\mu}+1) \tag{3b} \label{eq:3b}
\end{align}
\noindent
For notational convenience, we omit the position dependence in the last line and henceforth.

The scattering matrix $A_{\mu\mu'}$ can be separated into diagonal terms, representing decay terms (or inverse lifetimes $\tau_{\mu}^{-1}$), and off-diagonal terms $M_{\mu\mu'}$, constituting the mixing matrix.
Likewise, the phonon distribution $\psi$ may be expanded as:
\begin{align}
\psi&=\psi^{(0)}+\psi^{(1)}+\psi^{(2)} + \cdots, \tag{4} \label{eq:4}
\end{align}
\noindent
where $\psi^{(m)}$ collects contributions at $m^{th}$ order in $M$, encoding the population of carriers which are connected to the source $G_0$ by $m$ scattering events.
Decomposition of $A$ into decay and mixing terms and substitution of~\cref{eq:4} into~\cref{eq:2} returns the linearized recurrence relations
\begin{align*}
\left( \tau_{\mu}^{-1}+ v_\mu \cdot \nabla \right) \psi_\mu^{(m)} &= G_m(\mu)  \tag{5a} \label{eq:5a} \\
\intertext{and}
G_{m+1}(\mu) &= \sum_{\mu'} M_{\mu\mu'}\psi_{\mu'}^{(m)} 
\tag{5b} \label{eq:5b}
\end{align*}
These relations, indexed by scattering event $m$, illustrate our underlying assumptions of treating collisions purely in reciprocal space, and drift purely in real space, thus precluding any quantum effects in transport.
Each iteration in the algorithm represents a physical scattering event, the significance of which can be traced back to the thermalization of carriers, illustrated schematically in~\Cref{fig:1}.

Phonons are injected at a constant rate $G_0$.
They first drift in real space before scattering against the background phonon distribution via the mixing matrix.
This produces phonons at a different constant rate $G_1$, which is taken as the injection rate of phonons which have scattered once.
The procedure is then repeated until the convergence of~\cref{eq:4}, which is guaranteed by the positive-semidefinite nature of matrix $A$ (Supplementary Sections 1 and 2).

\begin{figure*}[ht]
\includegraphics[width=\linewidth]{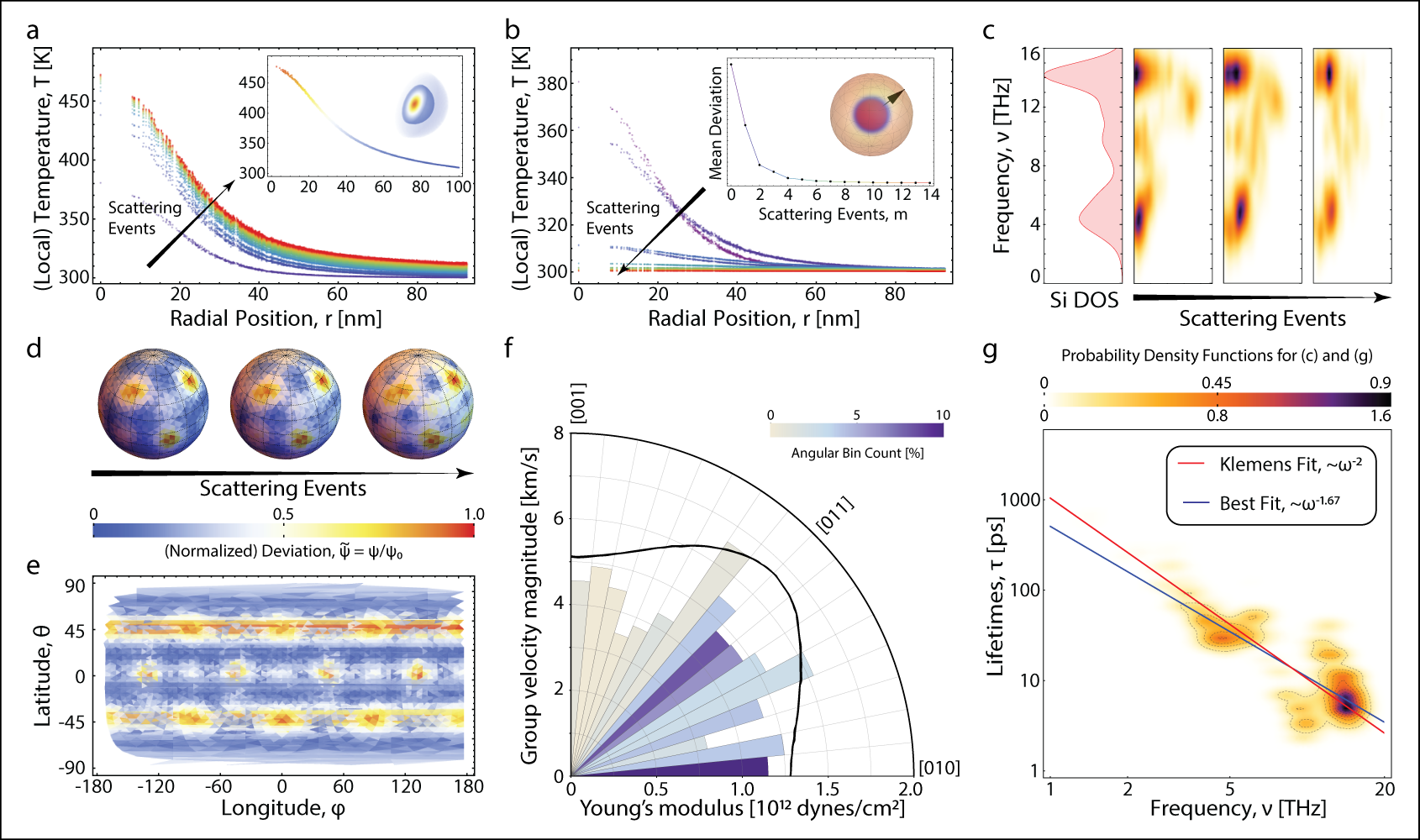}
\caption{Spatially-resolved thermal transport on silicon spherical nanoparticle. 
(a) Steady-state radial temperature profile for the first 15 scattering events.
Constant source term is a Gaussian profile (inset of (b)).
(b) Individual contributions to the temperature profile at order m, resembling the transient evolution of such an initial profile.
(c) `Heat' map of the distribution of carriers arriving at the surface of the nanoparticle as a function of scattering events.
(d) Accumulated surface distribution of carriers as a function of scattering events, highlighting the real-space anisotropy.
(e) Angular projection of surface distribution at first scattering event, illustrating accumulation at lines of constant latitude.
(f)  Phonon group velocity magnitudes as a function of direction. Black line draws the analogy with Young's modulus in the (100) plane, reproduced from~\cite{1965_Wortman}.
(g) Log-log plot of the \textit{ab initio} calculated phonon lifetimes using d3q~\cite{1998_Lazzeri,2013_Paulatto,2013_Fugallo,2014_Fugallo,2015_Cepellotti,2015_Paulatto}.}
\label[Fig]{fig:2}
\end{figure*}

To evaluate the utility of our spatially-resolved formalism we apply it to examine phonon transport in a silicon nanoparticle with a diameter of $200$ nm.
The steady-state distribution converges, in agreement with the prediction of the heat conduction equation, after a few scattering events (\cref{fig:2}(a)).
This is because the nanoscale dimensions of the nanoparticle favor ballistic rather than diffusive transport.

Despite operating in steady-state, our framework enables examination of individual scattering events, thus offering microscopic insight into transient processes inaccessible to the heat conduction equation. We find that scattering event-resolved distributions match the transient behavior obtained by treating the constant source term as an initial population instead (\cref{fig:2}(b)).
This highlights the physical origin of carrier thermalization as a direct consequence of scattering in the material and suggests that our formalism may offer insights into a variety of transient phenomena.
In many ways scattering events are a more physical descriptor of thermalization than time, which is effectively convolved with carriers' mean free paths.

Next, we analyze the accumulated distribution of carriers reaching the surface as a function of scattering events.
In momentum-space the `heat' maps suggest that carriers shift to higher energies as they scatter, with the acoustic phonon peak being absorbed by the optical phonon peak (\cref{fig:2}(c)).
In position-space, the distribution of carriers is anisotropic with `hot' and `cold' regions (\cref{fig:2}(d)).
This is a consequence of two phenomena; 
First, due to their ballistic behavior many carriers reach the surface without scattering. 
Second, the carriers exhibit anisotropic group velocities (\cref{fig:2}(f)) and consequently preferentially reach the surface at latitudes corresponding to densely sampled directions in the group velocity distribution (\cref{fig:2}(e)). This can be traced to the anisotropy of the elastic stiffness tensor's anisotropy for cubic materials, considering phonons in the long-wavelength limit~\cite{1965_Wortman}.
Subsequent scattering events work to isotropize the distribution (\cref{fig:2}(d)), a direct consequence of the mixing terms.

The spatial and scattering-event resolution of our formalism presents an opportunity to investigate interface transport outside the transmission-probability formalism~\cite{1995_Datta}.
When heat is conducted through interfaces the local temperature exhibits a sharp discontinuity, giving rise to Thermal Interface Resistance (TIR), which was first described by P. Kapitza in 1941 and has since been studied rigorously for a variety of materials~\cite{1969_Pollack}.

\begin{figure*}[ht]
\includegraphics[width=\linewidth]{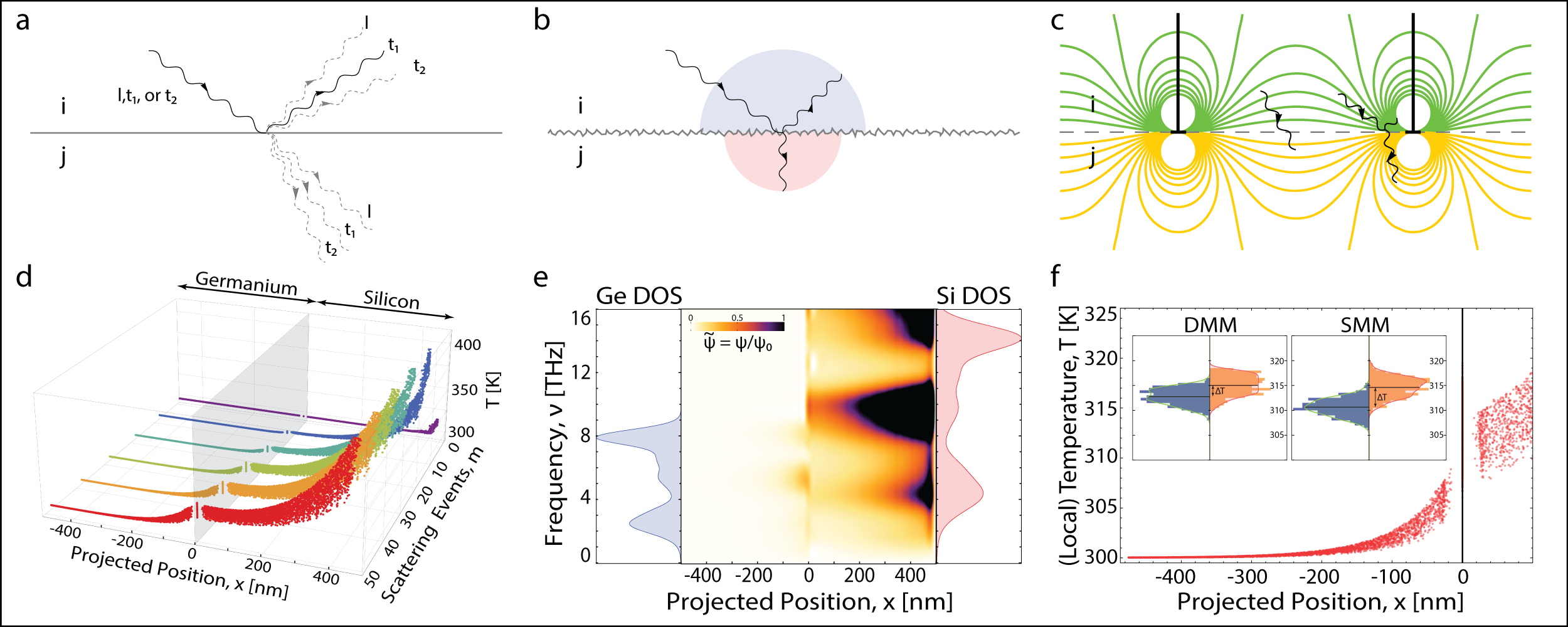}
\caption{Thermal interface resistance. 
(a) Diffuse Mismatch Model schematic.
The shaded blue and red semicircles illustrate the incoming phonon's `loss of memory' and diffuse scattering. 
(b) Acoustic Mismatch Model schematic. 
(c) Strain Mismatch Model schematic.
Green(Yellow) contours illustrate contours of constant compressive(tensile) dilatation.
(d) Steady-state temperature profile across a Si-Ge heterostructure using the DMM.
(e) Energy-binned phonon distribution across a Si-Ge heterostructure using the SMM.
(f) Direct comparison of DMM and SMM, highlighting the larger TIR for SMM.}
\label[Fig]{fig:3}
\end{figure*}

The earliest models to describe TIR, the Acoustic Mismatch Model (AMM)~\cite{1952_Khalatnikov}, and the Diffuse Mismatch Model (DMM)~\cite{1989_Swartz}, shown schematically in Figures~\ref{fig:3}(a) and~\ref{fig:3}(b), make use of the Landauer formalism attributing scattering to a mismatch of vibrational properties across the interface.
The AMM assumes planar interfaces where acoustic phonons can either specularly reflect, refract, or change phonon branch in an acoustic analogue of Snell's Law~\cite{1989_Swartz}.
The DMM replaces the complete specularity assumption with diffusive scattering at the interface, attributed to the atomically rough nature of the interface~\cite{1989_Swartz}.

While the influence of atomic roughness is especially significant at higher temperatures, both models underestimate TIR at moderate cryogenic temperatures and above, attributed to the omission of inelastic scattering~\cite{2005_Reddy,2001_Prasher}.
At higher temperatures, anharmonic interactions become important to TIR and despite the development of refined models to include full dispersion relations~\cite{2005_Reddy}, and address anharmonicity~\cite{2001_Prasher}), the corrections come at additional computational costs.
Recently, an alternative approach using non-equilibrium molecular dynamics (NEMD) has been proposed, capable of capturing TIR in the classical limit~\cite{2011_Termentzidis,2012_Termentzidis,2016_Gordiz,2017_Azizi}. 

We propose an alternative formalism in which we specify an interface Hamiltonian and compute transition probabilities using Fermi's Golden Rule (FGR).
Within our recursive framework, surface fluxes are expressed as
\begin{align*}
S_\mu=\left( \boldsymbol{v}_{\mu}\cdot \hat{\boldsymbol{a}}\right) \psi_{\mu}, \tag{6} \label{eq:6}
\end{align*}
where $\hat{\boldsymbol{a}}$ is the surface normal. This can be extended for an $i/j$ heterostructure as
\begin{align*}
S_\mu^{i,m+1} =  B_{\mu\mu'}^{i} \cdot S_{\mu'}^{i,m}+T_{\mu \mu'}^{i \rightarrow j} \cdot S_{\mu'}^{j',m}+ R_{\mu \mu'}^{i \rightarrow j} \cdot S_{\mu'}^{i',m} \tag{7} \label{eq:7}
\end{align*}
\noindent
and similarly for $S_\mu^{j,m+1}$. Here $B^{i}$ represents a surface bounce matrix, $T^{i \rightarrow j}(R^{i \rightarrow j})$ represents a momentum-resolved transmission(reflection) matrix, $m$ indexes scattering events, and the prime superscript specifies only those nodes shared at the interface.
This splits the surface phonon flux into material $i$ to a component being bounced at the surface of material $i$, a component being back-scattered at the interface from the incoming flux from material $i$ itself, and a component being transmitted at the interface from the incoming flux from material $j$.

Both the AMM and DMM can be reformulated with momentum-resolution for direct application and validation within our formalism. 
We illustrate the utility of our approach by proposing a structure-specific interface Hamiltonian for the case of semi-coherent interfaces such as a Silicon-Germanium heterostructure.
When two crystalline materials with similar lattice constants are brought in contact they form a semi-coherent interface and are characterized by the spontaneous formation of linear arrays of misfit (edge) dislocations along the interface~\cite{2007_Sutton}.
Although misfit dislocations are most-commonly found in epitaxially-grown interfaces, the model can be extended to all crystalline interfaces described by linear arrays of dislocations, such as low-angle grain boundaries~\cite{2007_Sutton}. 
The presence of the dislocations implies a periodic dilatational strain field along the interface, which couples with `bulk' phonons as first shown by Carruthers in 1959~\cite{1959_Carruthers}.

To derive the interface Hamiltonian describing phonon-strain coupling, we note that the potential energy of a crystal can be represented as a Taylor series around its equilibrium position in increasing powers of displacements, with the cubic term being the lowest anharmonic contribution~\cite{1990_Srivastava}.
In the presence of a strain field, one needs to consider the total atomic displacement, $u_{\rm{total}}=u_{\rm{ph}}+u_{\rm{strain}}$, where $u_{\rm{ph}}$ is the displacement due to phonon propagation and $u_{\rm{strain}}$ is the displacement due to the externally applied strain~\cite{1959_Carruthers}.
Expanding the cubic term we obtain four interaction categories: (1) thee-phonon interactions  ($u_{\rm{ph}}^3$); (2) two-phonon interactions with strain ($u_{\rm{strain}} u_{\rm{ph}}^2$); (3) one-phonon interactions with strain ($u_{\rm{strain}}^2 u_{\rm{ph}}$);  (4) vacuum interactions ($u_{\rm{strain}}^3$).
The first category is identified as regular three phonon interactions, found in the bulk independent of strain. 
The fourth category represents vacuum interactions which lead to a constant shift in energy and thus can be omitted.
Single phonon interactions with strain do not conserve energy to first order, and can be shown to cancel out exactly at higher orders~\cite{1959_Carruthers}.
We hence focus on two-phonon interactions with the strain field, described by the Hamiltonian:
\begin{equation}
 H'= \frac{\hbar}{4\rho \Omega} c_{2}(\mu_1,\mu_2) \prod _{i=1}^2 \left( a^{\dagger}_{\mu_{i}}+a_{\mu_{-i}} \right) \tag{8} \label{eq:8}
\end{equation}
\noindent
where $\rho$ is the material density, $\Omega$ is the unit cell volume, $a^{\dagger}(a)$ are the phonon creation(annihilation) operators, and $c_2$ is the phonon-strain coupling coefficient (Supplementary Section 3).
Applying FGR with initial and final states at either side of the heterojunction, we arrive at a conceptual picture of TIR: 
A phonon in state $\mu$ in material $i$ interacts with the Fourier component of the interfacial strain, scattering into a phonon in state $\mu'$ in material $i/j$, transferring the excess momentum to the strain field, allowing for symmetry breaking and inelastic scattering~\cite{1959_Carruthers}.
This model, termed the Strain Mismatch Model (SMM), is shown schematically in~\cref{fig:3}(c).
We use expressions for the dilatational strain field due to misfit dislocations from linear elasticity (Supplementary Section 3)~\cite{1996_Weertman}.

We compare SMM to DMM in a prototypical semi-coherent interface example of a heterojunction between $500$ nm cubes of Si and Ge. 
The local temperature discontinuity across the interface is recovered in position-space using DMM (\cref{fig:3}(d)).
Energy-binned carrier distributions using SMM, illustrate that phonon modes unable to be transmitted while conserving energy are scattered back into the material (\cref{fig:3}(e)). 
The `decay' lengths of the energy-binned carriers suggest a preference for higher energy carriers, consistent with findings for a single material nanoparticle (\cref{fig:2}(c)).
Comparing DMM to SMM at room temperature, by binning temperatures for either side and model, shows how SMM predicts a higher TIR than DMM (\cref{fig:3}(f)).

In this \emph{Letter}, we establish a theoretical and computational framework for semi-classical transport that describes all six degrees of freedom of the BTE at steady state. 
We have applied our recursive formalism to phonon transport, with \textit{ab initio} calculated scattering matrices, utilizing the multidimensional resolution to investigate the physical origins of phonon thermalization and anisotropic phonon distributions.
We have extended the framework to compute phonon surface fluxes and investigated heat transport across interfaces. 
Our perturbative formalism was validated against a semi-coherent interface within a Si-Ge heterostructure - a ubiquitous materials system in nanoelectronics.
The model confirms that non-intrinsic phonon scattering near the interface plays a dominant role in TIR, and provides a pathway for generalization to other structure-specific interfaces.
Our work may advance thermal transport engineering at nanostructured materials interfaces including those found in thermoelectrics, energy storage, and nanotheranostic agents.
By resolving individual scattering events, our formalism could also provide insight into transient behavior of phonons and capture non-equilibrium phenomena such as coherent phonon effects in all-optical characterizations of bandstructures of semiconductor heterostructures.

\section{Acknowledgements}
The authors thank Prof. Craig Carter and Nicholas Rivera of the Massachusetts Institute of Technology for fruitful discussions. GV and PN acknowledge funding from the Defense Advanced Research Projects Agency (DARPA) Defense Sciences Office (DSO) Driven and Nonequilibrium Quantum Systems program and the ONR grant on High-Tc Superconductivity at Oxide-Chalcogenide Interfaces (N00014-18-1-2691). ASJ thanks the UK Marshall Commission for financial support. ASJ acknowledges support by the Gordon and Betty Moore Fundation (GBMF7392) and the National Science Foundation (NSF PHY-1748958).

\end{document}